# A hidden chemical assembly mechanism: reconstruction-by-reconstruction cycle growth in HKUST-1 MOF layer synthesis

T. Koehler[a], J. Schmeink[b], M. Schleberger[b,c], F. Marlow[a,c]*

[a]Max-Planck-Institut für Kohlenforschung, Kaiser-Wilhelm-Platz 1, 45470 Mülheim an der Ruhr, Germany

[b]Fakultät für Physik, Universität Duisburg-Essen, Lotharstrasse 1, 47057 Duisburg, Germany

[c]Center for Nanointegration Duisburg-Essen, Carl-Benz-Str. 199, 47057 Duisburg, Germany

*Corresponding author (marlow@mpi-muelheim.mpg.de)

**ABSTRACT:** Thin metal-organic framework films grown in a layer-by-layer manner have been the subject of growing interest. Herein we investigate one of the most popular frameworks, type HKUST-1. Firstly, we show a synthesis procedure resulting in quick but optically perfect growth. This enables the synthesis of films of excellent optical quality within a short timeframe. Secondly and most importantly, we address the already known, but not fully understood observation that the expected monolayer growth is strongly exceeded in every single deposition cycle. This is an often-ignored contradiction in the literature. We offer a growth model using mid-cycle reconstruction process leading to a mathematically determined reconstruction-by-reconstruction (RbR) cycle growth with a 4-times higher growth rate representing an up-to-now hidden chemical assembly mechanism.

**INTRODUCTION**

Metal-organic frameworks (MOFs) are a well-known porous material class used for sensing, gas separation, catalysis, optics and many more applications. Their strength lies in the tunability due to the nearly infinite combination possibilities of different metal centers with organic ligands. Special interest has been attracted by thin surface-mounted MOF films (SURMOFs). This material class shows very promising characteristics for applications in optics,[1,2] sensing,[3,4] electronics,[5,6] as well as energy storage and conversion.[7] SURMOFs are typically synthesized by alternatingly exposing the substrate to the metal and linker reactants, with additional washing steps in between.[8,9] Usually this is reported to result in the growth of exactly one layer per cycle as shown in Figure 1, thus called layer-by-layer (LbL) synthesis. Recent reviews summarize this in principle epitaxial growth mechanism comprehensively.[7,10,11]

This work focuses on one of the most popular SURMOFs consisting of HKUST-1 or $Cu_2BTC_3$,[12] where 4-coordinate copper dimers are connected with 3-coordinate benzene-1,3,5-tricarboxylate (BTC) groups to form a 3D porous cubic network of $Fm\bar{3}m$ geometry via carboxylate-metal dimer bonds (data e.g. in[12,13]). For this network, LbL growth has been reported in a multitude of synthesis setups, such as dipping[9,14-16] or spraying[9,17-19] the substrate, as well as flow cells.[20] These sometimes require special and optimized equipment, thus providing a barrier for smaller groups to enter into the SURMOF field. Furthermore, immersion times in the order of 30 min to an hour per step are common. Such slow growth limits reasonably attainable film thickness *d*, whereas e.g.

optical applications would often require $d$ in the same order of magnitude as the wavelength of light. Recently, some progress had been made by reducing the total time of one synthesis to 8 h.[21] Here, we report synthesis conditions resulting in quick, but perfectly homogenous film growth on the macroscopic scale.

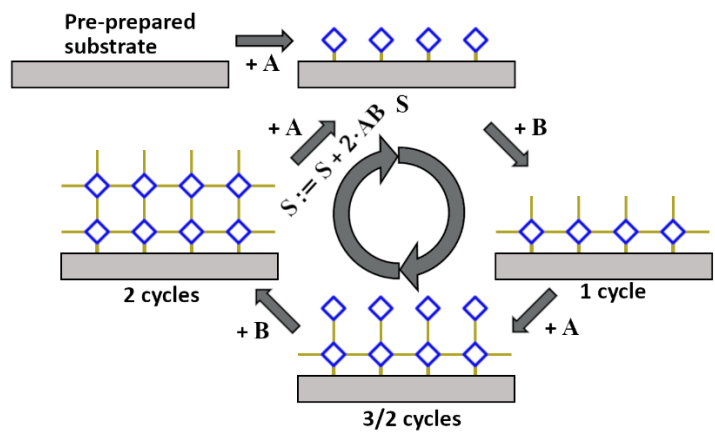

**Figure 1.** Standard layer-by-layer growth model[7,8,10,11] with alternating deposition of an $m$-coordinate metal center **A** ($m = 4$, blue) and 2-coordinate linker **B** (yellow) on a substrate. Intermediate cleaning steps (not shown) ensure the nearly complete removal of non-bonded reactants **A** and **B**, respectively. Only one layer per cycle $n$ is possible if the final structure forms immediately in every deposition step. Here, the number $m = 4$ is chosen because of display reasons, the mechanism also works for other values such as $m = 6$ which is more realistic for 3D growth. The assignment operation **S** := **S** + … is used for a mathematically description of the growth process.

The HKUST-1 cubic unit cell has a size of 2.63 nm, leading to a spacing of 0.65 nm between individual Cu-dimer layers in the (100)-direction (four per unit cell). The growth process, both as SURMOF and from solution, is often reported to take place in (111)-direction, which results in a layer spacing of 0.76 nm.[22,23] In the simple LbL model, one would therefore expect a thickness increase of either 0.65 nm or 0.76 nm per cycle $n$ depending on growth direction (see also the Supporting Information). However, this value is generally by far exceeded in the literature. Here, we address this discrepancy and provide a model using a reconstruction process in every cycle.

This model leads to understanding of the optimized synthesis procedure resulting in quick, but perfectly homogenous film growth on the macroscopic scale.

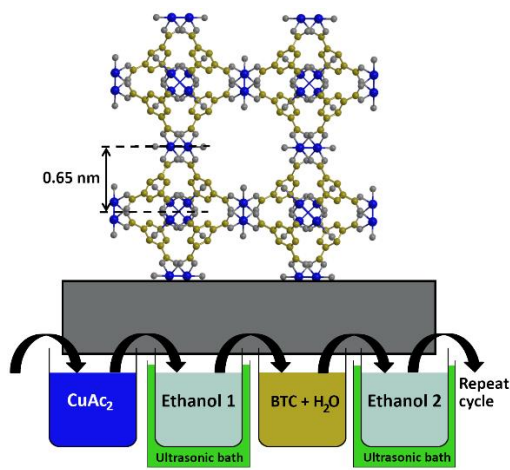

**Figure 2.** A possible HKUST-1 unit cell oriented with <100> in normal direction of the substrate and the applied dip-coating procedure with four immersion steps per cycle as used in this work. Atom colors: copper - blue, oxygen - gray and carbon - yellow. Lattice orientation and interface structure strongly depend on the synthesis procedure and are considered schematic here. A discussion of these issues can be found e.g. in Ref.[20].

## EXPERIMENTAL DETAILS

Since many of the HKUST-1-SURMOF synthesis procedures have proven to be difficult to be reproduced with exactly the same product appearance (film quality and defects), we report the used procedure in more detail than usual. Copper(II) acetate ($CuAc_2$) monohydrate (Alfa Aesar, 98.0 - 102.0 %) and benzene-1,3,5-tricarboxylic (BTC) acid (Sigma Aldrich, specified: 95 %, own analysis: 99.15 % of the organic part is BTC) were used without further purification. Absolute ethanol (99.5%) was purchased from Merck. MilliQ water for addition to the linker solution and substrate pre-treatment was purified in-house (18.2 MΩ at 22.0 °C). Glass slides (Thermo Scientific, Menzel glasses) were cut into 3 cm x 2.5 cm sized pieces and pre-treated by a two-step procedure: immersion for 30 min in a solution of Labosol lab-cleaning solution (5 % in milliQ water), followed by rinsing with deionized water. Immediately afterwards, the substrates were immersed for 30 min in a strongly basic solution (0.58 M NaOH in a 2:1 mixture of milliQ water and ethanol), followed by rinsing and drying in nitrogen. Finally, the substrates were treated in a plasma cleaner (PlasmaPrep 5, 5 W, 20 min).

Directly after this preparation step, two substrates were mounted on a sample holder back-to-back. The holder can be deposited on top of a beaker glass, resulting in 50 % immersion of the substrate into the reactant solutions. Two such holders were cycled through the reactant and cleaning solutions, resulting in alternating exposure of the substrates to the metal and linker solution. A full cycle consisted of the following steps (see also Figure 2):

1. $CuAc_2$ (30 s, 1 mM in ethanol)
2. Purging in ethanol (30 s, with ultrasonification)
3. BTC (30 s, 0.2 mM in 90 Vol% ethanol/10 Vol% $H_2O$)
4. Purging in ethanol (30 s, with ultrasonification, in a separate container to step 2)

During the transfer, the holder was carefully pulled out of the solution, then angled to about 45° and the excess liquid at the lowest corner of the substrate was stripped off by touching the inside of the beaker. This process takes about 7.5 s and results in a transfer of substrates that are still wet, but without large amounts of liquid being transferred between the four solutions. Nevertheless, all solutions were exchanged every 20 cycles. The solutions' level were chosen such that the substrate is immersed about 2 mm less into the reactant solutions than into the purging solutions. After the synthesis, ending with step 4 of the last cycle, the samples were left to dry in air and then held at 60 °C overnight to remove excess solvent molecules. The procedure uses parts of the ones given in[21,24-26].

The optical properties of the MOF films were analyzed with a UV/Vis spectrometer Cary 5G between (300 – 800) nm in double beam mode.

The X-ray diffraction data were collected on a Rigaku SmartLab equipped with a rotating anode (9 kW, 45 kV, 200 mA) in grazing incidence geometry (Cu $K_{\alpha1,2}$: 1.542 Å). Data were collected with a HyPix-3000 multi-dimensional detector in 1D mode. The samples were placed on a quartz sample holder and data were collected continuously in the range of 3° - 80° in steps of 0.01° and a scan speed of 0.5° min$^{-1}$. Before the measurement, sample height $z$ and tilting angle α are scanned and optimized. The incidence angle ω to the surface was optimized at the HKUST-1 peak with the maximum intensity at 2θ = 11.464° (222-peak) by maximizing the intensity at that specific angle as a function of ω.

Molecular vibrations were analyzed with Fourier Transform Infrared Spectrometry (FTIR) using a Magna 560 spectrometer in attenuated total reflection mode. For this, the sample was deposited upside down on a diamond crystal and pressed down gently with a stamp from the top until the thin film signal had been maximized. The signal intensity depends on the pressure of the stamp and the spectra were therefore normalized to the substrate glass peak at 904 cm$^{-1}$ during data processing.

AFM measurements were performed by a WITec alpha300 RA in tapping mode. The cantilevers used for measurement were NanoSensors' PPP NCHR extra sharp, reflective coated tips for tapping and non-contact mode. The cantilever driving frequency was set to 321 kHz, close to its resonance. The image resolution is 512 lines/image at 512 points/line and has a physical scale of 25*15 μm$^2$. To cut down on measurement time, fewer lines than 512 were sometimes collected. The topography was measured with a tip velocity of 1 line/s for forward and backward trace at a setpoint of 0.2 V for a 2 V driving amplitude. The P-I control was set to 10 % proportional and 8 % integral gain. These values were kept constant for all AFM measurements performed. The AFM images were processed and analyzed with *Gwyddion*.[27] First, the scanned rows were aligned (using the median of differences), followed by facet leveling. Finally, a mask was applied to the part of the image that belongs to the film and the whole image is plane levelled based on the masked area. The statistical analysis (height distributions, root mean square (RMS) roughness) was also based on the whole masked area.

**RESULTS**

HKUST-1 thin films are synthesized as detailed in the experimental section using very short immersion times of 30 s per step compared to previous reports (see the Supporting information). This is possible due to a combination of water as a modulator in the linker solution[24,25] (step 3 of the synthesis procedure) and ultrasonic bath treatment during the cleaning steps[26,28]. Ultrasonification during cleaning steps has been demonstrated to reduce surface roughness, potentially due to the detachment of large cluster form the surface.[28] Additionally, the presence of defects that absorb light in the visible range has been shown to be diminished.[26] These special synthesis conditions will be further discussed after analyzing the resulting MOF films. The use of water in other steps of the deposition method has not let to films with good optical quality.

The X-ray diffraction patterns (Figure 3) of the sample show good agreement to simulated curves, confirming successful HKUST-1 formation. The peak intensities indicate preferred growth in the 111-direction, as is often observed for HKUST-1 SURMOFs. However, other crystal planes are still present, pointing towards polycrystalline films. This is consistent with literature reports, which have concluded Volmer-Weber-like island growth instead of the Frank-Van-der-Merwe type deposition.[14,21] For our works,[19] polycrystallinity and lattice orientation is likely of lower importance because we seek for optical applications and a cubic system is optically isotropic. However, this may be different in other scenarios, such as those involving gas adsorption.

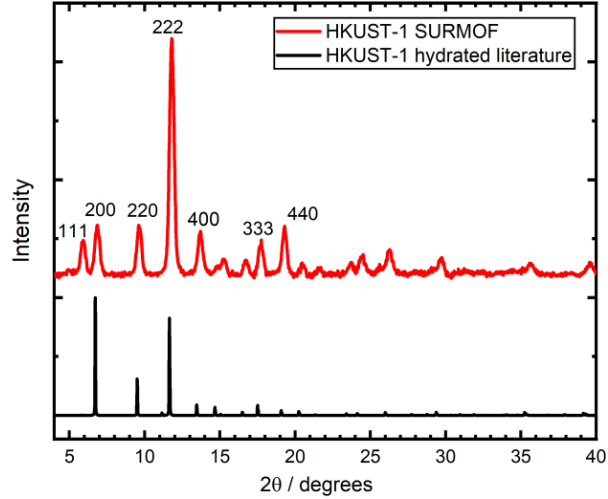

**Figure 3.** X-ray diffraction pattern of an HKUST-1 film (60 cycles) in grazing incidence geometry at ω = 0.22°, compared to simulated HKUST-1 pattern. [13] The peaks of the 111-family show an enhanced intensity compared to the other peak families.

Optical photographs and UV/Vis spectra (Figure 4a) confirm the excellent optical properties of the thin films. The visual appearance as depicted in the photograph provides no evidence of scattering, which justifies the designation of the film as optically perfect. A small absorption band around 700 nm can be attributed to a d-d transition of the copper and gives rise to a faint blueish color.[26] The measured extinction around 500 nm is even lower than that of the blank substrate. This can be explained by the fact that the measured extinction is entirely dominated by surface reflection in this range. The film acts as an anti-reflection coating, which is only possible if the refractive index is smaller than that of the substrate, in this case sodalime glass[29]:

$$n_{HKUST-1} < n_{glass} = 1.51 \tag{1}$$

This observation clashes slightly with the work of Redel *et al.*[18] who reported the refractive index of a hydrated HKUST-1 film to be between 1.55 and 1.6. A similar measurement of the film on quartz glass reveals a slight reflection enhancement and results in a lower boundary for the refractive index of 1.46 (see Supporting Information). Moreover, the UV/vis measurements in the transparent regions permit the estimation of scattering effects, which are below 0.3 %, i.e. lower than one tenth of the reflection losses. The results for other cycle numbers are very similar and yield the same limit for the scattering losses.

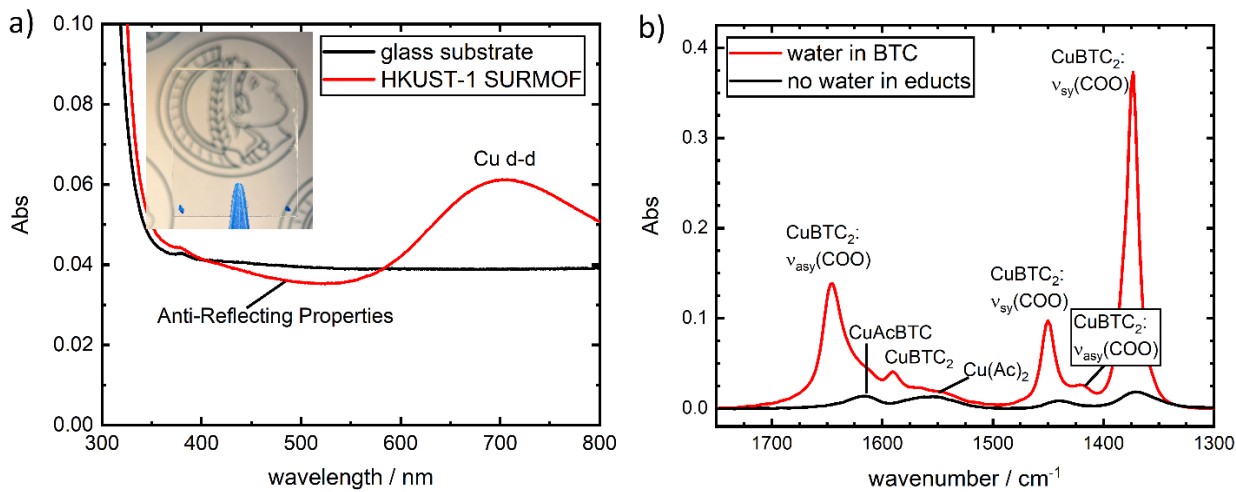

**Figure 4.** a) UV/visible extinction of an HKUST-1 SURMOF film with 80 cycles. Inset: Image of an 80-cycle sample under ambient irradiation. b) FTIR-spectra of an HKUST-1 SURMOF with water in the linker solution (red) and without it (black). The assignment of each peak to the kind of vibration is indicated (see text).

A key process during the synthesis is the exchange of the four acetate groups of the copper paddlewheels to the carboxylate groups of the BTC. This is an isomorphic exchange in a secondary building unit (SBU). The results can be well monitored by the molecular vibrations of the COO-groups using FTIR-spectrometry. The relevant part of the spectrum is shown in Figure 4b. We observe strong peaks at 1645 cm$^{-1}$, 1450 cm$^{-1}$ and 1373 cm$^{-1}$, as well as weak peaks at 1490 cm$^{-1}$ and 1420 cm$^{-1}$. Each of those is generally attributed to the HKUST-1 framework, though some discussion on the exact vibrational assignment remains.[21,30,31] Delen et al.[21] furthermore identified additional peaks at 1620 cm$^{-1}$ and 1550 cm$^{-1}$ and assigned them to non- or only partially exchanged paddlewheels. These peaks are observed only very weakly in our spectra, confirming the excellent chemical quality resulting from our synthesis method. The same can be said for the peak belonging to dangling acid bonds at 1705 cm$^{-1}$, which also shows the very low defect density of our samples.

Figure 4b also shows the FTIR spectra of the films synthesized without the addition of water in the linker solution. The difference is striking. The previously mentioned HKUST-1 peaks are only barely observed, while non- or partially exchanged paddlewheels make up the majority of the signal. This shows clearly that water plays a crucial role during the synthesis. In fact, the effect is even stronger in our case than in the literature reports first introducing the beneficial effect of water in the linker solution.[24,25] Very likely, the reason for that is the short immersion time of only 30 s per step. In pure ethanol, 30 s is not enough time for a mediator-free exchange of the acetate to the linker group and, therefore, most LbL HKUST-1 reports in the literature utilize immersion times upwards of 30 min per step to achieve a significant deposition. We think that water as synthesis modulator in the BTC step is an alternative and more efficient solution to this problem.

Furthermore, the reproducibility of the UV/Vis and IR spectra is noteworthy. As illustrated in the Supporting Information, the measured spectra were nearly identical for four different samples. The dependence on cycle number was found to be nearly linear for the lattice-related peaks and constant for the defect peaks.

Water improves the quality of the sample in terms of defects, macroscopic film properties, and sometimes orientation, as also mentioned before.[24,25] Nevertheless, the biggest benefit seems to be that HKUST-1 SURMOF can be synthesized with a speed of 2 min/cycle, while still retaining good optical, structural, and chemical quality. This eases access into the field of SURMOF by groups, which do not have access to sophisticated, automated setups. Furthermore, it makes reaching a higher cycle number and thereby a higher thickness *d* feasible in a laboratory environment. Higher sample numbers are also possible, as will be demonstrated in the next paragraphs.

To validate the theoretical predictions of the standard growth mechanism, as outlined in the introduction, we have conducted a detailed AFM study. Key questions to answer were whether $d$ truly increases linearly with $n$ and if so, whether the increase during one cycle is equal to a single layer only. Fifteen samples were synthesized for this purpose. A scratch was made down to the substrate to facilitate a clean edge for measuring the step height in the AFM image.

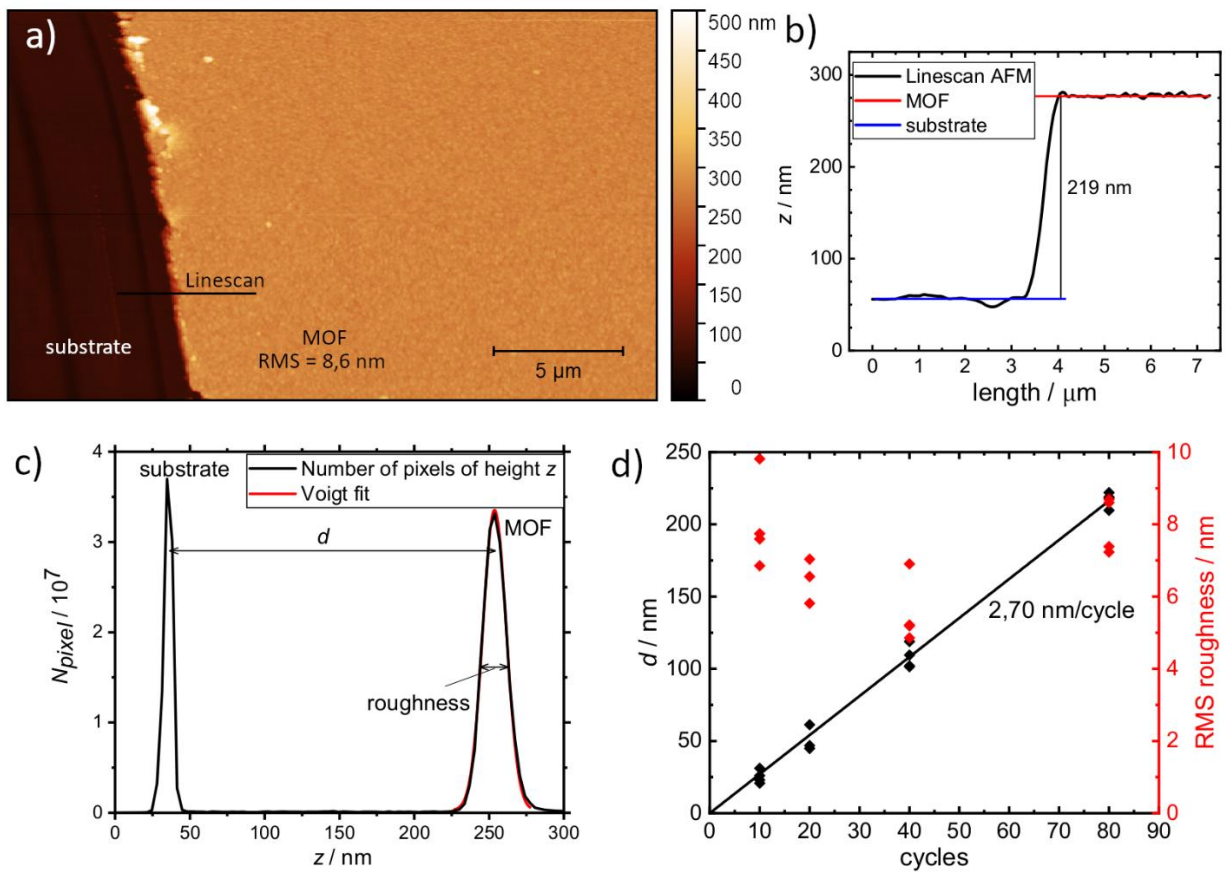

**Figure 5.** a) AFM image of an 80 cycle sample with a scratch down to the substrate on the left side and b) line scan (width 1 pixel, linear interpolation) over the edge. c) Number of pixels $N_{pixel}$ measured at a given height $z$ (black). The peak belonging to the MOF surface is fitted with a Voigt peak function (red). d) The thickness $d$ as determined from the height distributions and RMS roughness of all measured samples as a function of cycle number $n$.

One of the SURMOF films grown in 80 cycles is shown in Figure 5; all other AFM images can be found in the Supporting Information (Figures S2-S5). A line scan (Figure 5b) shows a sharp jump in height over the edge of the scratch. For each image, the height distribution of all pixels was evaluated, as in Figure 5c. We observe two distinct peaks: one belongs to the MOF surface and one to the substrate. $d$ can be determined from the distance between these 2 peaks and agrees well with the values determined from line scans (see Supporting Information Table S2). To avoid the partial arbitrariness of the line selection, we have chosen $d$ determined from height-probability diagrams for our further evaluation. The results of our cycle number dependent thickness study are plotted in Figure 5d as a function of $n$. It can be clearly seen that the film thickness does indeed increase linearly with $n$. However, the slope is $(2.70 \pm 0.03)$ nm/cycle, which needs to be further discussed. This finding is consistent with ref.[20], which analysed the growth rate using a quartz crystal microbalance for several systems. In this pioneering paper, it was observed that the growth exhibited linear behaviour, and a number of growth models were developed based on this observation.

In addition, no indication of induction cycles was discerned in our experiments, i.e. no slower or faster growth rate for low cycle numbers. This finding also agrees with ref.[20], where only the BTC step of the first cycle differs visibly from the other cycles. Nevertheless, the induction period is of significant importance for many MOF growth processes[32,33] and should be studied in more detail in future works.[32,33]

The mean roughness of the films is about 6 nm - 8 nm. No dependence on $n$ was found. Together with the cycle-independent thickness increase per cycle, it seems that we have no accumulation of film errors with $n$ up to 80 cycles. This could point to an absence of single layer defects or to a self-healing process during growth.

**GROWTH MECHANISM**

The highly constant thickness increase of 2.70 nm/cycle clearly contradicts the simple LbL mechanism (with 0.65 nm/cycle to 0.76 nm/cycle[22,23]) and instead indicates the growth of four layers per cycle. Scouring the literature, we found that we were not the only ones to observe this discrepancy.[14-16,18,20] Interestingly, the thickness discrepancy exists for very different LbL synthesis procedures. In fact, there are only very few works reporting a proper measurement series $d(n)$ of HKUST-1 SURMOF with monolayer growth rate; one exception is ref.[34]. However, the majority of measurement results in the literature clearly contradict the simple growth model (Figure 1) that is very popular in the MOF literature.

One possible explanation for the four times higher growth rate could be cross-contamination between the reactant solutions. For practical reasons, we have only exchanged solutions every 20 cycles and we have transferred the sample without complete drying. This could lead to small amounts of reactants being transferred between the containers, facilitating larger than monolayer growth. More specifically, the reactants would need to first be transferred into a cleaning solution and from there into the second reactant solution. Only when both copper and linker are present in one of the solutions at the same time, one would expect a deviation from the layer-by-layer growth as shown in Figure 1. To quantify the effects of cross-contamination, we have analyzed the purging solution after the BTC immersion step (solution 4) for presence of BTC using high-performance liquid chromatography (HPLC). We found a BTC-concentration of 1.34 µM (See Fig. S6-S8), which corresponds to about 19.0 pmol of the linker being transferred into the copper solution per cycle even at the most contaminated state of the purging solution right before its exchange. This is less than 1 % of the substance amount necessary to form a single monolayer (full calculation in the Supporting Information). Cross-contamination of the solution can, therefore, be clearly ruled out as the cause of the 4-fold increase in growth rates observed in our case.

Having excluded the standard LbL growth model and possible contamination effects as a reason for deviations, we would like to propose a modified growth mechanism. While this is an attempt to solve the contradiction, which we have pointed out, we must emphasize that it is simply one logically consistent possibility and may require further modification. Our proposed growth model is based on two basic assumptions: (i) every deposition leads to a surface with a maximum molecular density for either reactant **A** or **B** and (ii) a reconstruction can take place, but only in the presence of water or after long reaction times. The reconstruction results in a structure, where all inner bonds are saturated. The resulting growth model is shown in Figure 6. Some simplifications had to be made for displaying the more complex 3D structure in 2D and Figure 6 displays the topological relations only. It describes the following stages and steps:

**Stage ①**: The surface is covered with copper dimers, which are in turn saturated with the 3-coordinate BTC, together forming parts of the well-known paddlewheels. In the ideal monocrystalline structure, many of these would already have multiple connections to Cu dimer, but at this stage, they do not find free reaction sides.

**Stage ②**: The reactant **A** has been added, saturating the surface with a maximum number of copper paddlewheels. More Cu has been added than needed for the growth of a single layer due to additional reactive sites available at the outer surface.

**Stage ③**: Reactant **B** has been added together with water. The water facilitates an efficient reconstruction of the excess Cu paddlewheels together with the provided BTC. The step ends with a reconstructed HKUST-1 lattice forming stage ③.

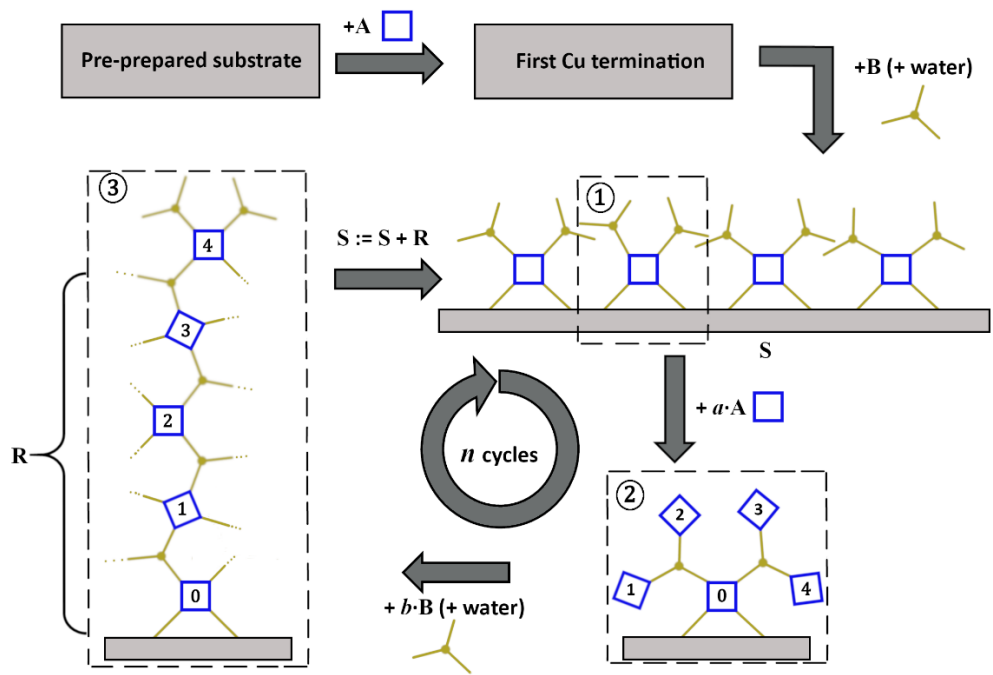

**Figure 6.** Schematic description of an L₄bL₄ growth process consisting of the interplay of surface enrichment and lattice reconstruction. Color code: 4-coordinate Cu dimers **A** in blue, 3-coordinate BTC-linker **B** in yellow. Stages: ① 2**B**-enriched surface, ② 4**A**-enriched surface, ③ reconstruction sheet R with a 2**B**-enriched surface. The three dots indicate incorporation into the lattice by saturated bonds.

This would result in $R = 4L = 4(AB_{4/3})$ and the following uptake parameters per step and per Cu-dimer in the surface layer:

$$a = 4 \;,\; b = 16/3 \tag{2}$$

The structure of the reconstruction sheet R is considered to be saturated and equal to HKUST-1. A ligand exchange in the SBUs is taking place during the chemical reactions between stage ① and ② as well as between stage ② and ③.

Generalizing the model for α-coordinate A notes and β-coordinate B notes delivers a modified reconstruction sheet $R = aL = a(AB_{\alpha/\beta})$. The uptake parameters can be found by counting:

$$a = \frac{\alpha}{2}(\beta - 1) \;,\; b = a\frac{\alpha}{\beta} \tag{3}$$

Of course, for every generalized system it is necessary to prove that a final structure with saturated bonds does indeed exist.

In the aforementioned growth model, the framework reconstruction between stages 2 and 3 is of key importance. This involves the breaking and re-formation of bonds between Cu paddlewheels and BTC. Water is known as efficient modulator for the exchange of acetate to carboxylic acid.[24,25] We believe that it plays a similarly crucial role in the formation of intermittent steps during the suggested reconstruction processes. The reconstruction likely occurs during very long immersion times as well and this might be the reason that several research groups have gravitated towards immersion times upwards of 30 min (60 times longer than here). However, often the involvement of trace water cannot be fully excluded and its role might be underestimated in previous reports.

In light of the proposed growth model, one may better understand the findings of previous literature. Stavila et al.[20] observed during in-situ mass uptake measurements of flow cell synthesis of HKUST-1, that more mass gets absorbed during the first BTC addition than suggested by the stoichiometry. Afterwards, the stoichiometry follows the expected 3 Cu to 2 BTC ratio well, albeit at an overall increased rate similar to the one observed in this work. These findings agree well with our assumption (i) and support the proposed model.

**CONCLUSION**

Experimental conditions resulting in quick, but optically perfect HKUST-1 films were found. Exceptional chemical purity of the framework was shown using FTIR spectroscopy. The applicability as optical thin films was established using UV/Vis spectroscopy. Important aspects of the synthesis are the utilization of water as a modulator in the BTC ligand solution, as well as ultrasonic bath treatment steps in between. This allowed for a crucial reduction in immersion times, enabling manual synthesis of high cycle numbers, and thereby, high film thicknesses within a few hours.

An AFM investigation of samples grown with different cycle numbers revealed a four times larger than monolayer thickness increase per cycle. This is in clear contradiction to the frequently used layer-by-layer model, which is rooted deeply in the SURMOF literature and concepts. We offer a possible modification to the growth model. In this model, much more material than needed for one monolayer can be taken up per cycle and redistributed under the right experimental conditions via a reconstruction process occurring in every deposition cycle. This can feasibly explain both the observed strong influence of water during the ligand addition, as well as the 4-fold increased growth rate. The reconstruction process modifies therefore the overall picture of the growth: in addition to alternating deposition, reconstruction-by-reconstruction (RbR) become the key steps of the growth. We also believe this concept could be applicable to other SURMOFs with a linker coordination number larger than two.

The growth via the RbR mechanism represents a variant of self-assembly assignable in-between deposition of molecular monolayers and spontaneous generation of complete 3D structures. A modulator, in our case water, can be used to initiate a single reconstruction step in every cycle and therefore controls the whole process. Since the process strongly differs from LbL growth, one can also expect different defect generation and healing processes. The thickness-independent roughness found in this work points towards this direction.

## ACKNOWLEDGEMENTS

The authors gratefully acknowledge DFG for funding (projects Ma 1745/9-2 and 429784087), Jan Ternieden for X-ray measurements, the HPLC department at KOFO for HPLC measurements,

ASSOCIATED CONTENT

**Supporting Information**

The Supporting Information is available free of charge at https://pubs.acs.org/doi/10.1021/acsmaterialsau.XXXXX. The HKUST-1 lattice and layer structure, synthesis times, details of the atomic force microscopy, HPLC of the purging solutions, calculation of the effect of cross-contamination, an estimation of scattering effects, the reproducibility of the samples are described there.